%% file: NP_excitons_paper9.tex
\documentclass[aps,prb,twocolumn,showpacs,superscriptaddress]{revtex4}  
\usepackage{graphicx}  
\usepackage{dcolumn}   
\usepackage{bm}        
\usepackage{amssymb}   
\usepackage{multirow}
\usepackage{amsmath}
\usepackage{natbib}

\begin{document}

\title{Tight-binding calculations of image charge effects in colloidal nanoscale platelets of CdSe}

\input Author_list.tex

\date{\today}
\begin{abstract}
CdSe nanoplatelets show perfectly quantized thicknesses of few monolayers. They present a situation of extreme, yet well defined quantum confinement. Due to large dielectric contrast between the semiconductor and its ligand environment, interaction between carriers and their dielectric images strongly renormalize bare single particle states. We discuss the electronic properties of this original system in an advanced tight-binding model, and show that Coulomb interactions, including self-energy corrections and enhanced electron-hole interaction, lead to exciton binding energies up to several hundred meVs.
\end{abstract}
\maketitle

\section{Introduction}
Colloidal nanoplatelets (NP) are atomically-flat, few monolayers-thick semiconductor nanostructures\cite{Ithurria2008}. They are produced in a highly controlled manner, using the soft chemistry techniques of colloidal nanocrystal growth\cite{Bouet2013}. So far, II-VI semiconductors like CdSe\cite{Joo2006, Ithurria2011a}, CdS\cite{Li2012, Son2012} and CdTe\cite{Ithurria2011b} have been investigated. Nanoplatelets grow in the Zinc-Blende phase, with a [001] axis, and are terminated by Cd planes on both sides, which implies a significant non-stoichiometry: a n-monolayer NP consists of n planes of Se and n+1 planes of Cd. These nanoplatelets form thanks to the saturation of Cd dangling bonds on (001) surfaces by organic ligand molecules, which block the growth in the [001] direction while the platelet extends along other crystallographic directions in the layer plane. The detailed mechanisms are still under discussion, but clearly involve the bonding of carboxylic acid to Cd. Importantly, this passivation prevents the Fermi level pinning into mid-gap surface states, and associated non-radiative recombination paths. As a matter of fact, NPs show very promising optical properties with strong and narrow emission lines at both cryogenic and room temperatures\cite{Tessier2012}.
Ensembles of billion NPs show ground exciton optical linewidths as small as 40 meV, for quantum confinement energies of the order of 1 eV. This indicates that thickness fluctuations are well below a monolayer. Actual thicknesses and flatness were recently assessed by high-resolution on-edge TEM images\cite{Malher2012}. From a modeling perspective, these new nano-objects are ideal to test electronic structure theories in a regime of extreme, yet perfectly defined quantum confinement. Clearly, for thicknesses in the 1-2 nm range, only large-scale first-principle calculations or atomistic modeling within the atomistic pseudo-potential or the tight binding frameworks can provide a quantitative account of single particle states. However, one must also consider a strong renormalization of bare electron and hole states by the \textquotedblleft dielectric confinement\textquotedblright\cite{Rytova1965, Keldysh1979} effect due to proximity of the ligand/solvent with a smaller dielectric constant. Carriers in the semiconductor induce a surface polarization that is classically accounted for by introducing virtual \textquotedblleft dielectric image\textquotedblright charges.  Repulsive interactions between carriers and their own images produce self-energy contributions that increase dramatically the bandgap when the semiconductor layer thickness decreases to the nm scale. Conversely, when real electron and hole come close to each other as in the exciton ground state, each carrier interacts not only with its partner, but also with an infinite set of partner image charges, which substantially increases the electron-hole interaction\cite{Muljarov1995}. In such systems, the exciton optical transition energy results from conflicting effects of electron and hole self-energies and exciton binding energy enhancement\cite{Gippius1998} due to increased electron-hole interaction. Here, we combine advanced tight-binding calculations of single particle states and effective mass description of in-plane dispersion to calculate excitonic properties of semiconductor nanoplatelets.

\section{Bare electron and hole states}
The extended-basis $spds*$ tight binding model \cite{Jancu1998} is known as an efficient empirical-parameter full-band representation of semiconductor electronic properties. Parameter transferability from bulk to quantum heterostructures is very good. Of special importance is the model ability to represent vacuum states using ``vacuum atoms" that can be used in the tight-binding formalism and account for the vacuum/semiconductor interface \cite{Jancu1998, Jancu2008}. However the model has inherent parameter richness and its major difficulty is the solution of the ``inverse problem" of finding TB parameters out of known features of the bulk band structure. This was done systematically for III-V and IV-IV semiconductors, but not yet for II-VI materials. We start with a parameterization of bulk CdSe obtained from an interpolation of \textit{ab initio} electronic structure in the LDA+GW approximation and experimental band gaps in  both wurtzite (WZ) and zinc-blende (ZB) phases. TB parameters are listed in Table I, significant features of electronic structure are compared with available experimental data in Table II, and corresponding band structure is shown in Fig. 1. 

{\small
\begin{table}[h!]
\caption{Tight-binding parameters used in calculations.}
\begin{tabular}{cr|cr}
\hline 
\multicolumn{4}{c}{Parameters for CdSe (eV)}\\
\hline
$              a$ & $    6.0520$ & & \\
$         E_{sa}$ & $   -9.0819$ & $         E_{sc}$ & $    4.3707$ \\
$       E_{s^*a}$ & $   17.0529$ & $       E_{s^*c}$ & $   17.0896$ \\
$         E_{pa}$ & $    2.1891$ & $         E_{pc}$ & $    7.7150$ \\
$         E_{da}$ & $   13.9804$ & $         E_{dc}$ & $   14.0348$ \\
$       ss\sigma$ & $   -1.3225$ & $   s^*s^*\sigma$ & $   -2.5343$ \\
$ s_cs^*_a\sigma$ & $   -2.0814$ & $ s_as^*_c\sigma$ & $   -1.4142$ \\
$   s_ap_c\sigma$ & $    2.1639$ & $   s_cp_a\sigma$ & $    2.7620$ \\
$ s^*_ap_c\sigma$ & $    2.4498$ & $ s^*_cp_a\sigma$ & $    2.0491$ \\
$   s_ad_c\sigma$ & $   -2.5519$ & $   s_cd_a\sigma$ & $   -1.7505$ \\
$ s^*_ad_c\sigma$ & $   -0.9572$ & $ s^*_cd_a\sigma$ & $   -0.9549$ \\
$       pp\sigma$ & $    3.7728$ & $          pp\pi$ & $   -0.8875$ \\
$   p_ad_c\sigma$ & $   -1.4494$ & $   p_cd_a\sigma$ & $   -0.9185$ \\
$      p_ad_c\pi$ & $    1.3840$ & $      p_cd_a\pi$ & $    1.2855$ \\
$       dd\sigma$ & $   -0.9523$ & $          dd\pi$ & $    1.9375$ \\
$       dd\delta$ & $   -1.6454$ &                  &      \\
$\Delta_a/3$ & $    0.1656$       &  $   \Delta_c/3$ & $    0.0809$ \\
\hline
\end{tabular}
\end{table}
}

\begin{figure}[b!]
  \includegraphics[width=\linewidth]{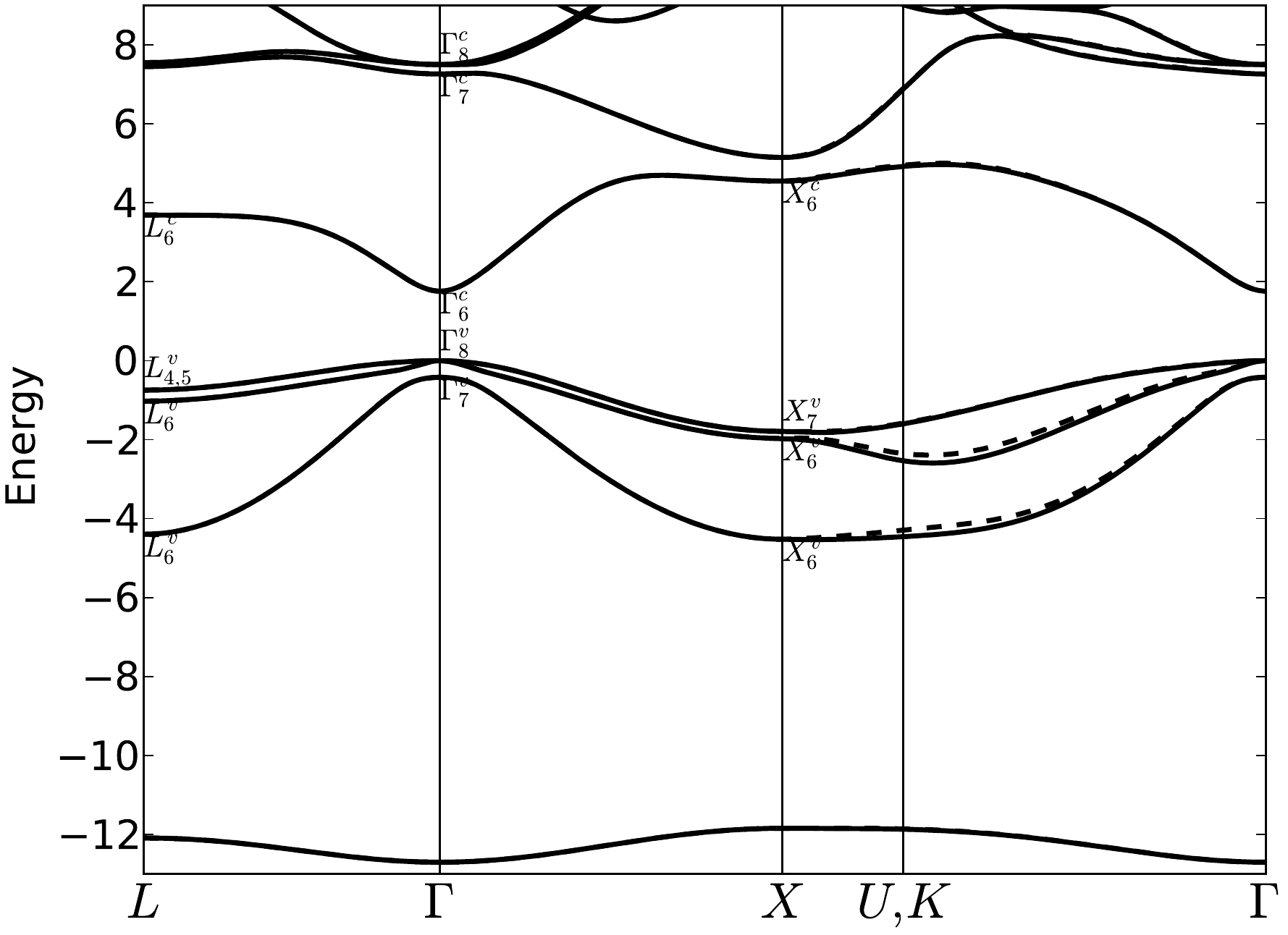}
  \caption{Calculated ZB-CdSe band structure\cite{notewv}}
  \label{f:dispersion}
  \end{figure}

\begin{table}[t]
\caption{Some calculated ZB-CdSe band parameters compared 
with available experimental data.}
\begin{tabular}{ccc}
\hline
 & TB & Expt. \\
\hline
$E_g$                 & $1.76$ & $1.76$\footnotemark[1],$1.66$\footnotemark[2], $1.74$\footnotemark[3] \\
$\Delta_{so}$         & $0.42$ &                        $0.41$\footnotemark[2]\footnotemark[3] \\
$m_e$                 & $0.104$& $0.12$\footnotemark[2] \\
$E(L_{4,5v}-L_{6c})$  & $4.43$ & $4.314$\footnotemark[2], $4.28$\footnotemark[3] \\ 
$E(L_{6v}-L_{6c})$    & $4.72$ & $4.568$\footnotemark[2], $4.48$ \footnotemark[3]\\ 
$E(X_{7v}-X_{6c})$    & $6.34$ & $~6.0$\footnotemark[2] \\ 
$\gamma_1$    & $4.243$ & \\
$\gamma_2$    & $1.415$ & \\
$\gamma_3$    & $1.801$ & \\
\hline
\footnotetext[1]{PRB 50, 8012 (1994)}
\footnotetext[2]{PRB 49, 7262 (1994)}
\footnotetext[3]{JAP 78, 4681 (1995)}
\end{tabular}
\end{table}

CdSe band structure has some specificities that are worth mentioning: compared to III-Vs, the Cd and Se $s$ and $p$ orbitals are much deeper, resulting in a larger energy separation with the quasi-free electron states $d$ and $s*$  and a lesser influence of the latter on L and X conduction minima. A most important consequence is the low optical index resulting from the correspondingly large value of E1 and E2 bandgaps. This is a general feature of II-VI compounds as compared with otherwise similar III-V's. Present parameterization yields a low frequency value of the dielectric constant $\epsilon_r^\infty=4.6$, somewhat smaller but in fair agreement with the reported experimental value $\epsilon_r^\infty=6$. However, for small energies, further screening by optical phonons occurs, and the experimental value of exciton binding energy in bulk CdSe, 10 meV, actually corresponds to a dielectric constant closer to the static dielectric constant $\epsilon_r^0=10$.

\par
\begin{figure}[t]
  \includegraphics[width=\linewidth]{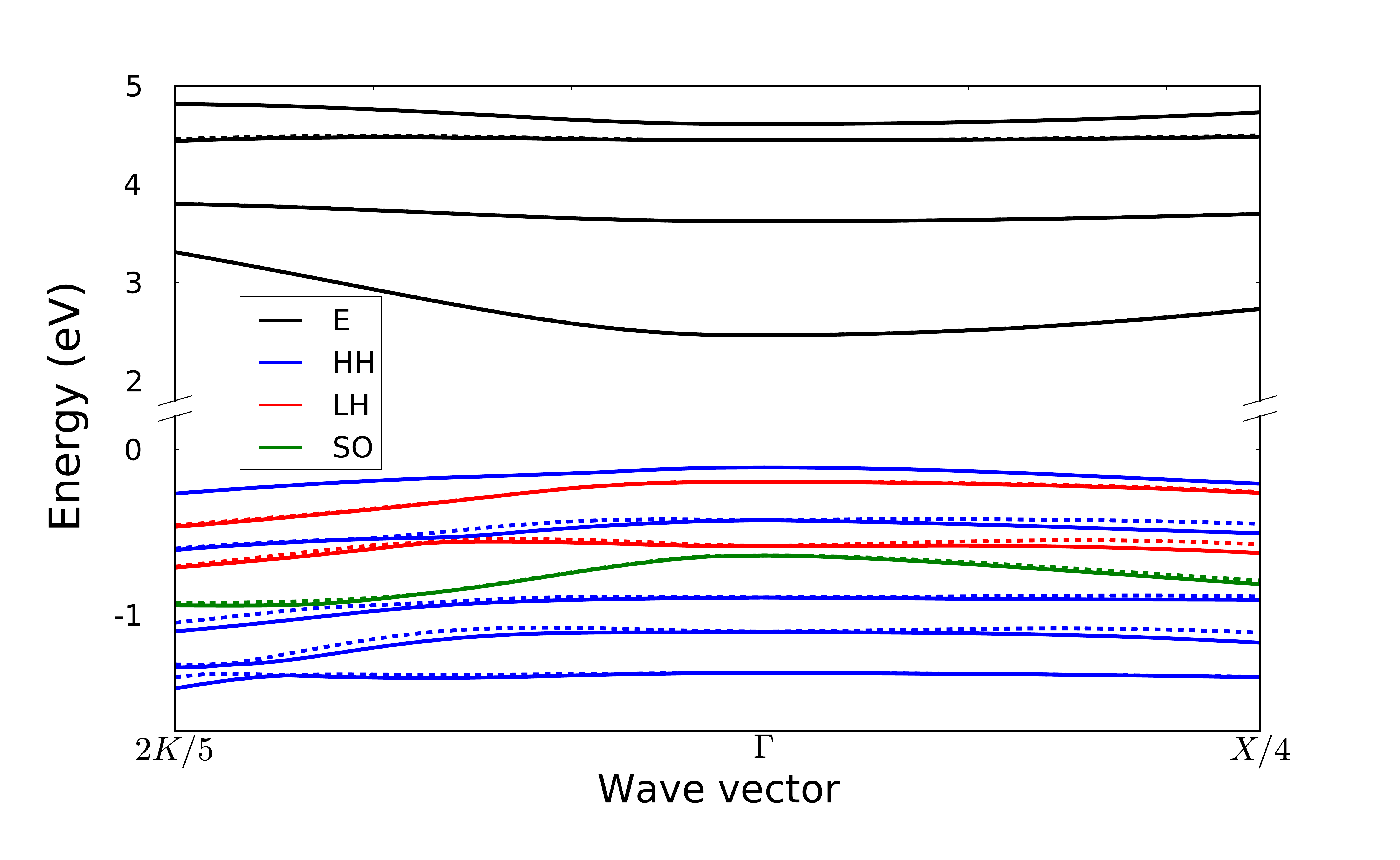}
  \includegraphics[width=\linewidth]{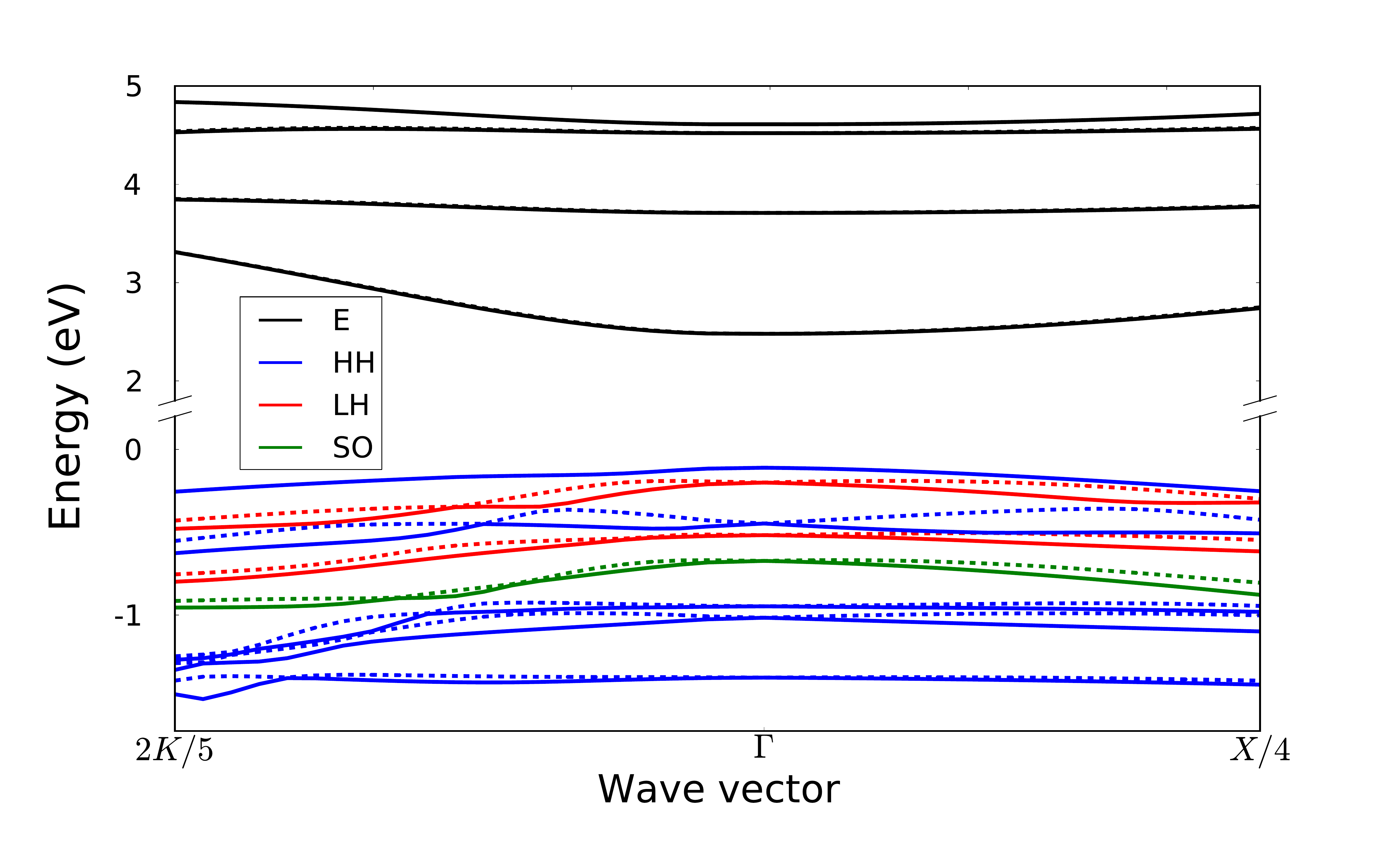}
  \caption{in-plane dispersion of bare single particle states, for a NP thickness of 5 monolayers, Cd (top) vs Se (bottom) termination. The two spin states for each level are shown with solid and dashed lines. K and X refer to the $\pi/a (1,1,0)$ and $\pi/a(1,0,0)$ wavevectors. The origin of energies is the bulk valence band maximum.}
  \label{f:disp}
\end{figure}

 Next, using this TB parameterization, we model the electronic properties of nanoplatelets with either Cd or Se terminations, and surrounded by suitable ``ligands atoms". Note that an atomistic method is definitely required to account for nanoplatelet stoichiometry defect. It is well known that clean, real $(001)$ surfaces show mid-gap pinning of the Fermi level due to dangling bonds. We insist that midgap states are not an artefact of the tight-binding method, they correspond to physical reality for clean surfaces and are actually found in first principle calculations. In particular, imperfect passivation of surface states is generally considered as responsible for the ``blinking" properties of colloidal nanoparticles. Conversely, if these dangling bonds are transformed into covalent bonds (in tight-binding modeling, hydrogenation is a standard trick for that), midgap states disappear and ``vacuum", or actually, ``ligands", act as a large barrier in both conduction and valence bands. The way carboxylic ligands attach to CdSe nanocrystals is a topic of interest in quantum chemistry \cite{Voznyy,Koposov}. Due to perfect 2D translational invariance, NPs present an original situation, highly favorable to modeling with first principle methods. However, to the best of our knowledge, such studies have not been reported yet. In absence of a detailed description of the ligand/semiconductor interface, we adopt here a simplified approach of tuning the parameters of the ligand/semiconductor interface in such a way that they simulate a large barrier, with a 20 eV bandgap, and valence (resp. conduction) band offsets equal to 7.7 (resp. 10.6) eV. In these conditions, calculated electron and hole states are insensitive to details of ``ligand" parameters. Main results for Cd-terminated NPs are summarized in Table \ref{t:table1}, and the in-plane dispersion for a 5 monolayer ($ml$) NP is shown in Fig. \ref{f:disp} for hypothetic Se-terminated and actual Cd-terminated NPs.

\begin{table}[b]
\caption{Main properties of ground electron ($E_1$) and hole ($H_1, L_1,SO_1$) single particle states, where H, L  and SO  stand for heavy, light and split-off bands. $E_{conf}$ (in $meV$) is the bare confinement energy. $m^\star$ is the in-plane effective mass (in $m_0$ unit). The bandgap, spin-orbit splitting and electron effective mass of ZB CdSe are $E_g^0 = 1.761 eV$, $\Delta_0= 0.42 eV$ and $m_e^{\star} = 0.104 m_0$.}

\begin{center}
\begin{tabular}{llccccc}
\hline
 \multicolumn{2}{c}{$Thickness$}  & 3   & 4   & 5   &6       &7   \\
\hline
$E_1$  & $E_{conf}$               &1333 &  955&  719& 562    &  451   \\

       &$m^\star/m_0$             & 0.35& 0.27& 0.22&  0.19  &  0.17 \\
\hline
$H_1$  & $E_{conf}$               & 209 & 147 & 110 &  86    & 69  \\

  &$m^{\star(100)}/m_0$           & 0.52& 0.45& 0.41&  0.38  & 0.35 \\
\hline
$L_1$  & $E_{conf}$               & 304 & 241 & 200 &  169   & 145  \\

  &$m^{\star(100)}/m_0$           & 0.55& 0.52& 0.50&  0.50   & 0.52 \\
\hline

$SO_1$  & $E_{conf}+ \Delta_0$              & 1010& 789 & 673 & 604   & 559  \\

  &$m^{\star(100)}/m_0$           & 0.52& 0.38& 0.33&  0.34 &  0.43  \\
\hline

\label{t:table1}
\end{tabular}
\end{center}
\end{table}

\begin{figure}[b]
  \includegraphics[width=\linewidth]{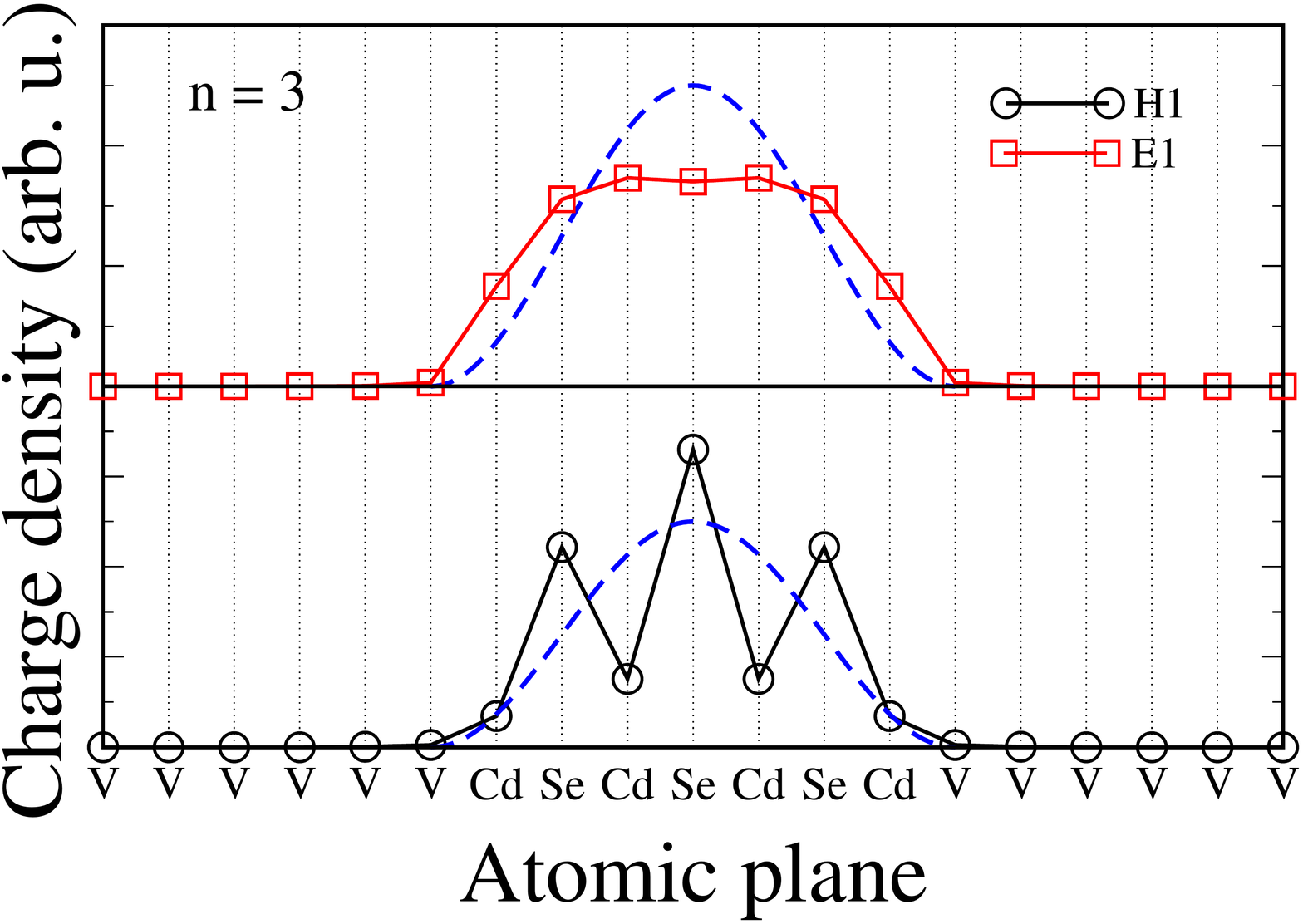} \includegraphics[width=\linewidth]{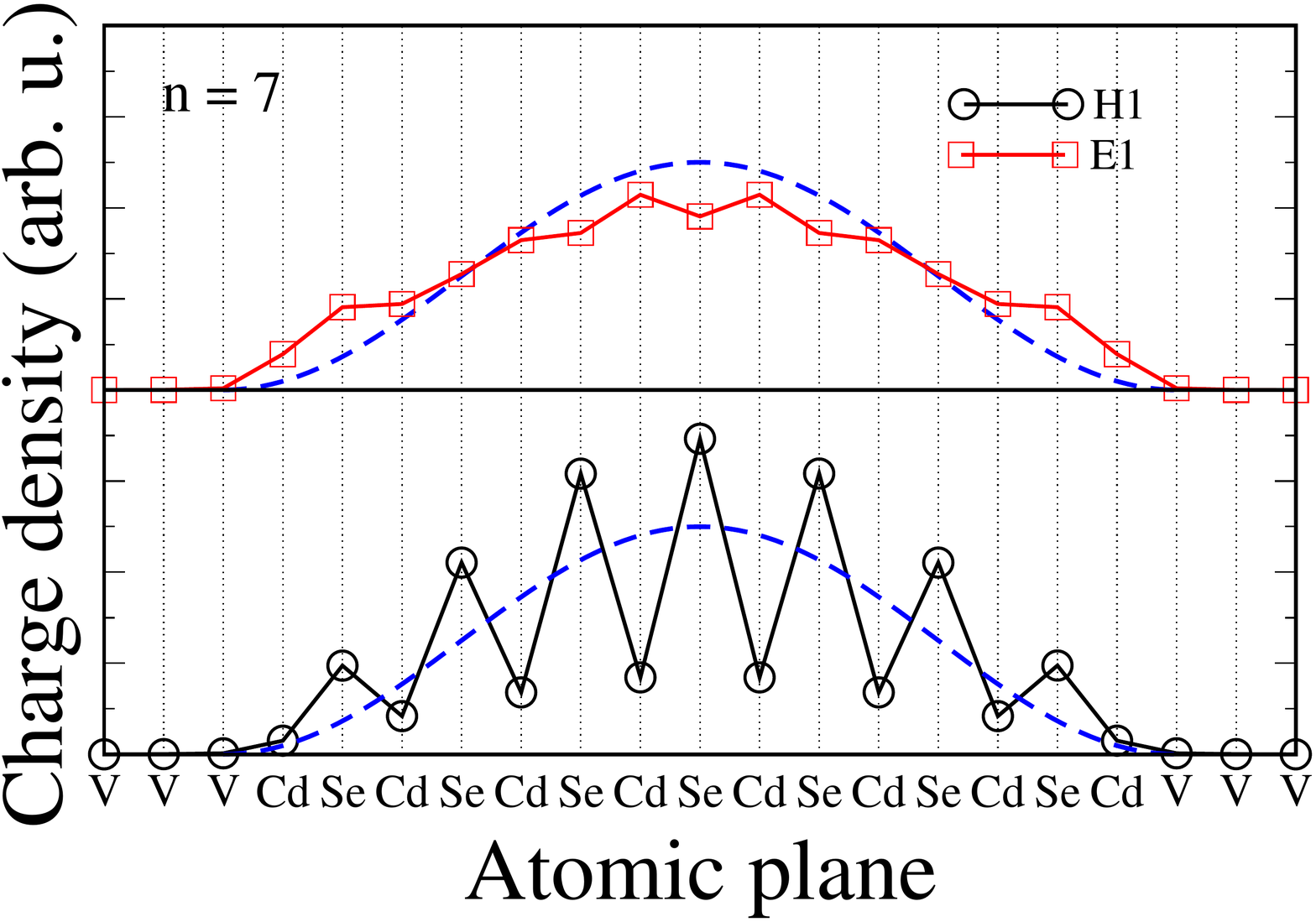}
  \caption{Electron and hole TB amplitudes for 3 (top) and 7 (bottom) monolayer thick, Cd-terminated NPs. The squared sinewave charge distribution for an infinite well effective mass model with thickness $L= (n+1/2) ml$ is also shown for comparison (dashed lines)}
  \label{f:amplitudes}
\end{figure}

Electron quantum confinement reaches the 1 eV range. Yet it is much smaller than the naive evaluation $\hbar^2\pi^2/2m_e^{\star}L^2$ (where L is the NP thickness and $m_e^{\star}$ the band-edge electron effective mass), due to strong non-parabolicity effect. Non-parabolicity also manifests itself in the in-plane dispersion showing a conduction band effective mass increasing strongly with decreasing thickness, and reaching up to three time the bulk band-edge mass. As for valence subbands, quantum confinement is comparable to spin-orbit coupling and eigenenergies appear in the energy range of the inflection points of bulk band structure. For this reason, the number (resp. spacing) of valence subbands is considerably larger (resp. smaller) than what would be expected from the consideration of bulk band-edge masses. In Fig.\ref{f:amplitudes}, we show the plane-averaged tight-binding amplitudes for the ground electron and heavy hole states for various NP thicknesses. The envelope of these amplitudes departs quite significantly from a sinewave, even if one smoothes the expected anion vs cation amplitude difference. This behavior of envelopes reflects the fundamentally multi-band character of electron and hole states in such extreme confinement regime. The comparison of Cd-terminated and Se-terminated NPs in Fig. 1 shows that while quantum confinement itself does not depend much on stoichiometry defect, spin splittings, in particular in the valence band, are very sensitive to exact composition. Finally, the bare single particle states obtained in the present tight-binding calculations differ appreciably from previous 8-band k.p results\cite{Ithurria2011a}, in spite of similar band edge parameters ; for instance, for the 5 monolayer  NP, k.p calculations of ref. \cite{Ithurria2011a, Ithurria_thesis} predicted ground electron and heavy hole confinements E1= 777 meV, H1= 161 meV. Present results are more reliable, due to much better representation of bulk valence and conduction dispersions for large k-values in the TB representation.

\section{Dielectric confinement}

\par Yet, this simple view of bare single particle states must be corrected for self-energy effects due to interaction of electrons or holes with the surface polarization that they themselve induce in order to fulfill electrostatic field continuity relations across the dielectric interface between the NP and its ligand/solvent environment. This problem is elegantly treated using the theory of ``dielectric image charges". A carrier in the large dielectric constant medium undergoes repulsive interaction with its image charges. These self-energy effects become large in ultrathin layers like NPs, since (in a continuous media approach), the repulsive self-interaction potential diverges when a carrier approaches the dielectric interface\cite{Kumagai1989, Muljarov1995, Gippius1998}. In an effective mass model, no divergency occurs even if the dielectric constant undergoes an abrupt change at any position, because the charge density associated with envelope functions remains finite. Physical reality is more complex, as dielectric function builds-up on a length scale of the order of 1-2 bond lengths, and charge is distributed in a wavefunction that can be represented as the product of a rapidly varying microscopic wavefunction by the slowly varying envelope function. For this reason, atomistic modeling of the self-energy is difficult: both the microscopic charge distribution and the position and profile of the dielectric interface directly come into play in a very sensitive way. First-principle calculations can give reasonably  accurate values for both quantities \cite{PhysRevB.74.045318}, but they usually have limited precision for bare single particle states, and have high computational cost. Here, the planar-averaged tight-binding charge densities for electrons and holes (see section II) are used to calculate the self-energies within an approximated scheme: we consider an abrupt jump of the dielectric constant at some distance $\delta$ of the last Cd atoms plane.  This simple approach mimics a well-localized microscopic function (eg Gaussian) with a half-width of the order of $\delta$. With $\delta$ in the $0.1-0.2 nm$ range, this is a sensible approximation to actual microscopic charge distribution. This scheme can eventually be improved by considering realistic local wavefunctions \cite{Benchamekh2013}. We treat this calculation to first order in perturbation (i.e. we do not recalculate single particle states in the self-energy potential). This could easily be improved by implementing a self-consistency loop, however such refinement is not important in regard of uncertainties on dielectric constants and simplifications in the model: indeed, second order perturbation would not couple the ground state to the first excited state, but only to more distant states of even parity. In Table IV, we compare results obtained using the static dielectric constant (experimental value $\epsilon_r^0 = 10$), valid for carriers with low kinetic energy, with those obtained using the high frequency dielectric constant  $\epsilon_r^\infty = 6$, valid in the limit of kinetic energies larger than optical phonon frequencies. Note that in NPs quantum confinement exceeds by far optical phonon energies, so $\epsilon_r^\infty$ is definitely more relevant in this problem. The ligand/solvent dielectric constant is taken equal to 2. For the $\delta$ parameter we retain a value of $0.1 nm$; increasing $\delta$ up to $0.2 nm$  decreases calculated self-energies by a typical 20$\%$.

\begin{table}
\caption{Ground electron and heavy-hole self-energies $E_{self}$ (in meV), calculated using the corresponding tight-binding charge distributions (see Fig. 4), $\epsilon_{ext}=2$ and  for $\epsilon_{NP}$  either the static $\epsilon_r^0=10$ or high-frequency $\epsilon_r^\infty=6$ values of the dielectric constant.}
\begin{center}
\begin{tabular}{llccccc}
\hline
 \multicolumn{2}{c}{$Thickness$}  & 3 & 4 &5 &6 &7\\
 \hline
 $E_1$&
$\epsilon_r^0=10$    & 186   & 148   & 122   &104   & 90  \\
&$\epsilon_r^\infty=6$   & 205   &163  & 135   &114   & 99  \\
\hline
$H_1$  &
$\epsilon_r^0=10$       & 173    &  138 & 114   & 97  & 84  \\
&$\epsilon_r^\infty=6$     & 188   &150  & 124   &106   & 92  \\
\hline
\label{t:table2}
\end{tabular}
\end{center}
\end{table}

\begin{figure}
  \includegraphics[width=\linewidth]{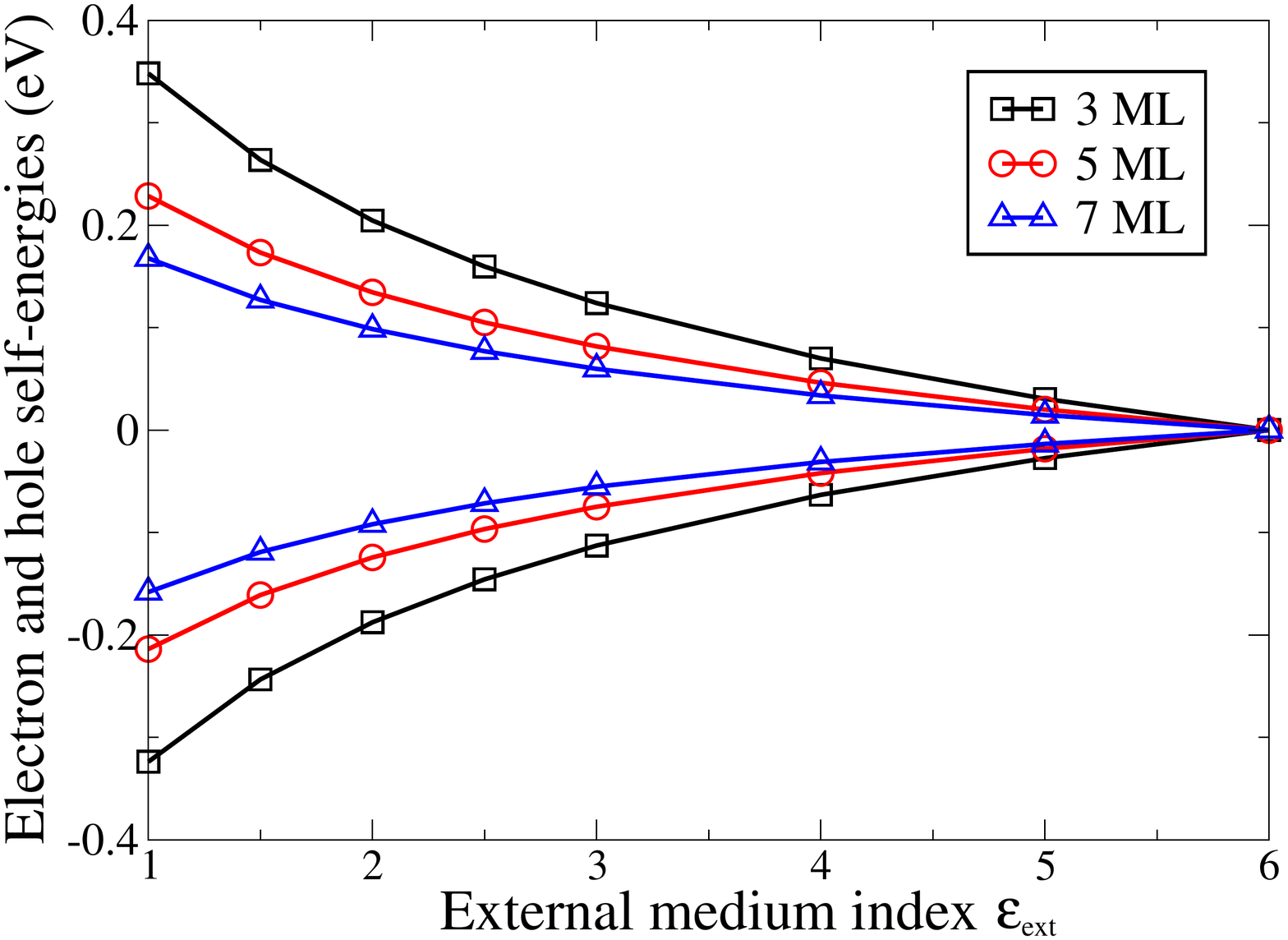}
  \caption{single particle self energies as a function of $\epsilon_{ext}$,  for $\epsilon_{NP} = 6$ and selected NP thicknesses of 3 and 7 monolayers.}
  \label{f:self-energies}
\end{figure}

\par Electron and hole self-energies (see Table IV) sum up and increase the NP bandgap $E_g^{NP}= E_g^0 + E_{conf} + E_{self}$ quite significantly. In Fig. \ref{f:self-energies}, the variation of electron and heavy-hole self-energies as a function of external medium index $\epsilon_{ext}$ is shown for  $\epsilon_{NP} = \epsilon_r^\infty = 6$ and NP thicknesses of 3, 5, and 7 monolayers. Electron and heavy-hole self-energies differ slightly because charge distributions differ, but the relative difference is quite small. Ground light-hole and split-off hole also display similar values of self-energies. Conversely, excited levels have larger self-energies because the corresponding charge distribution is on average closer to the dielectric interface. 

\par
\begin{figure}
  \includegraphics[width=\linewidth]{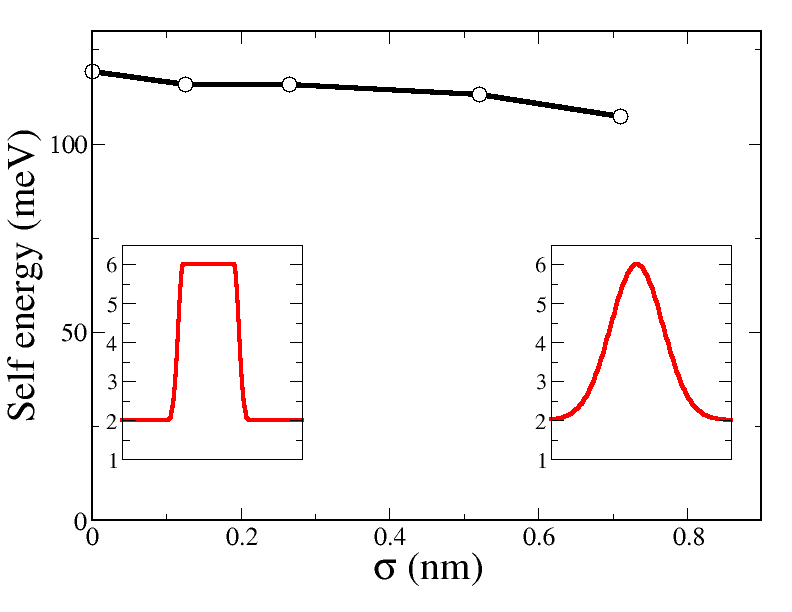}
  \caption{Self-energy calculated in an effective mass model for  a $1.67 nm$ (= $5.5 ml$)-thick platelet, with gradual change of dielectric constant.
  The insets show the dielectric constant profiles for $\sigma = 0.125 nm$ (left) and $\sigma = 0.71 nm$ (right).}
  \label{f:self-energies1}
\end{figure}

 Finally, it is interesting to check the effect of a smooth instead of abrupt change in dielectric constant. Indeed, it is known that dielectric constant builds-up on a length scale comparable to 1-2 bond lengths, which is not vanishingly small compared to NP thickness. For this evaluation, we used the simple effective mass approach with infinite potential barrier, and considered an arbitrary dielectric constant profile consisting of a plateau terminated with half-gaussians. Results, shown in Fig. \ref{f:self-energies1} for the case of a $1.67 nm$ (= $5.5 ml$)-thick platelet, indicate that self-energies are remarkably insensitive to the abruptness of dielectric constant profile, until the width of the plateau vanishes. Note also that the magnitude of self-energy using a squared-sinewave charge distribution and $\epsilon_{ext}=2$,  $\epsilon_{NP} = 6$, $E_{self} =110 meV$,  compares favorably with corresponding values using tight-binding amplitudes for the 5 monolayer NP.

\section{Excitons}

\begin{figure}
  \includegraphics[width=\linewidth]{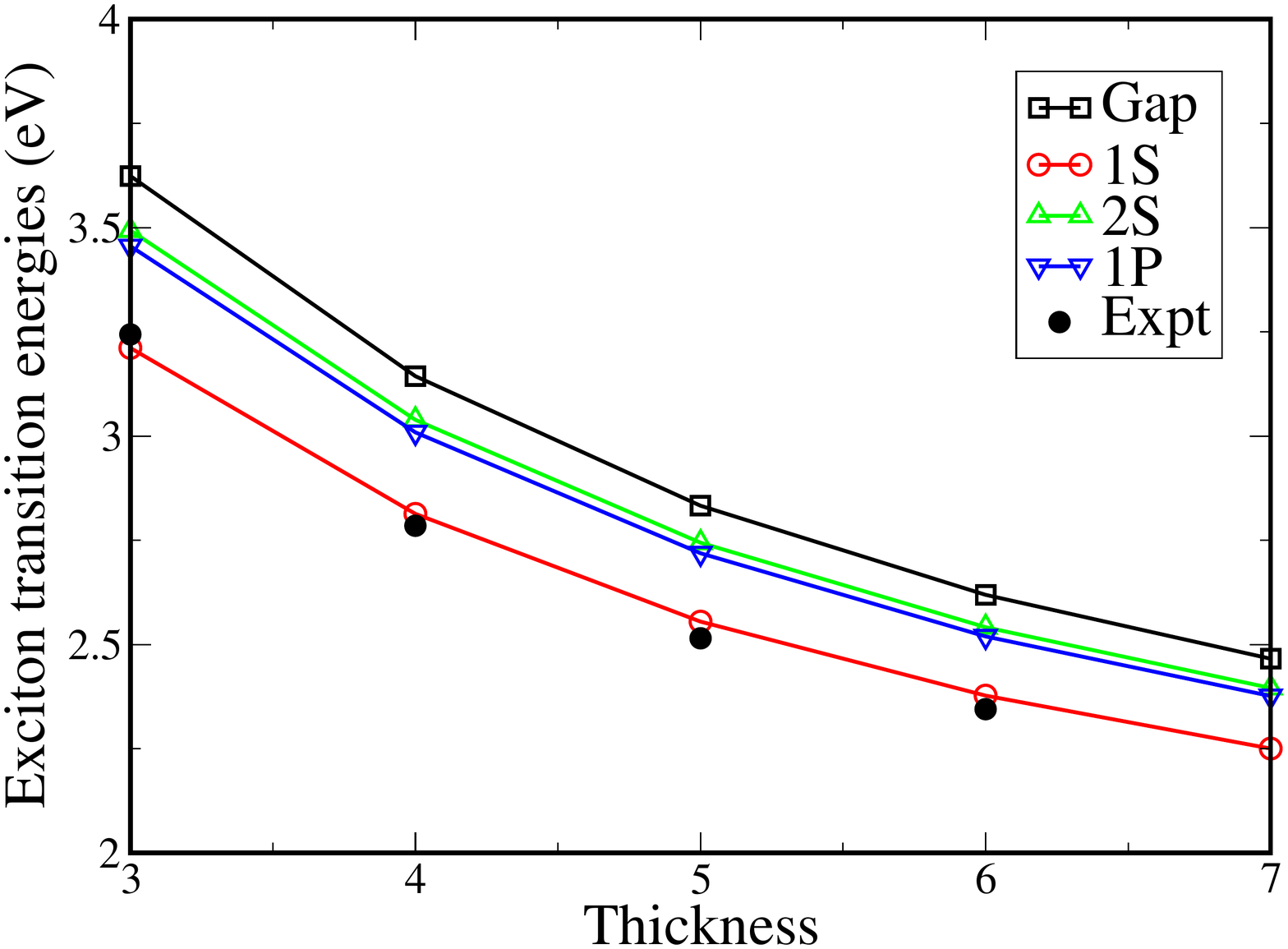}
  \caption{H1-E1 exciton transitions energies $1S, 2S, 1P$ and $\infty S$ (=$E_g^{NP}$) versus thickness, using $\epsilon_{ext}=2$, and $\epsilon_{NP} = 6$. Experimental data points from ref. \cite{Ithurria2011a} are also shown.}
  \label{f:exciton}
\end{figure}

Large in-plane effective mass, strong 2D confinement and increase of electron-hole interaction due to image-charge effects obviously combine and produce strong exciton binding energies. In principle, the method of full configuration interactions could be used together with tight-binding eigenstates of a finite lateral size platelet to fully model Coulomb interaction \cite{Bester2003}. This computationally difficult approach is far beyond the scope of the present contribution. Here, we adopt a much simpler scheme using the TB amplitudes for the wavefunctions along the z direction together with an effective mass approach for the in-plane motion, using the TB effective masses (see Table III). Since we just aim at evaluating the binding energy of an electron-hole pair we restrict ourselves to the main, direct term of Coulomb interaction and neglect electron-hole exchange. This approach is similar to the classical one\cite{Kumagai1989, Mosko1997, Gippius1998}, with the exception of using tight-binding amplitudes instead of envelope functions for the axial wavefunctions $\psi_{e,h}(z)$. In a first calculation, we use the experimental bulk value $\epsilon_r^0 = 10$ for the dielectric constant. Calculated binding energies for various NP thicknesses are shown in Table V. The remarkably large enhancement over the CdSe bulk Rydberg (10 meV) is actually governed by three factors: $i)$ the large (\textgreater 2) enhancement of electron effective mass; $ii)$ the dimensionality reduction and $iii)$ the electron-hole image-charge interactions. In order to isolate the dimensionality effect, we calculate the 3D Rydberg using the electron and hole in-plane effective masses, and compare it with the exciton binding energies in absence of image-charge effects. We thus find that the enhancement due to reduced dimensionality is fairly constant for the investigated thicknesses, being in the range 2.5$\sim 2.7$. The largest contribution is the image charge effect. For the exciton ground state, the attractive effect of electron and hole image-charges partly compensates the dominant repulsive effect of single particle self-energies: the excitonic transition energy is slightly $\it{larger}$ than, but remains close to the bare single particle bandgap $E_g^0 +E_{conf}$: the ground transition energy is in fact not strongly affected by Coulomb interaction \cite{Kumagai1989, Gippius1998}. However, for excited states $nS$, the electron-hole interaction decreases and the effect of self-energies prevails more and more as $n$ increases, so that the $nS$ transitions are strongly blueshifted with respect to the bare single particle gap. Finally, we point that the calculated binding energies are much larger than optical phonon energies in CdSe ($\sim 20 meV$). This implies that a dielectric constant close to $\epsilon_r^\infty = 6$ should be used to calculate NP exciton ground state \cite{Bajaj2003}. Hence, CdSe displays the remarkable property that changing the layer thickness allows a continuous tuning between weak and strong excitons, with binding energies respectively smaller and (much) larger than optical phonon energies. Note that a theory allowing the interpolation between the "weak" and "strong" exciton regimes has already been developed \cite{Zheng1998a, Zheng1998b, Bajaj2003}, but its implementation in the present context appears unnecessary, since all involved states have kinetic energies much larger than LO-phonon energies. Results corresponding to $ \epsilon_{NP} =\epsilon_r^\infty = 6$ are also given in Table V, and evidence even larger binding energies. In Fig. \ref{f:exciton}, we show the variation of exciton transition energies $1S, 2S, 1P$, and  $\infty S$ (=gap) versus thickness, for $\epsilon_{ext}=2,  \epsilon_{NP} = 6$, together with experimental results taken from ref. \cite{Ithurria2011a}. We used room temperature absorption spectra and added  95 meV to account for the temperature dependence that was actually measured in luminescence.\\

\begin{table}[h!]
\caption{H1-E1 exciton binding energies (in $meV$) for different NP nominal thicknesses and dielectric constants $\epsilon_{NP}$. We take $\epsilon_{ext}=2$.}
\begin{center}
\begin{tabular}{rlccccc}
\hline
\multicolumn{2}{c}{$Thickness$}  & 3 & 4 & 5 & 6 & 7\\
\hline
\textit{without images}    & $\epsilon_r^0 = 10$   & 71 & 58 & 50  & 44  & 40  \\
\textit{without images}    & $\epsilon_r^\infty = 6$   & 168  & 136  & 116  & 103  & 93  \\
\textit{including images} & $\epsilon_r^0 = 10$ & 289  & 231 & 193  & 168  & 149  \\
\textit{including images}  & $\epsilon_r^\infty = 6$   & 413  & 330  & 278  & 242  & 216  \\
\hline
\label{t:table3}
\end{tabular}
\end{center}
\end{table}

The agreement is fairly good, but might be a little bit fortuitous, due to existing uncertainties in several important parameters. There is indeed room for deepening our understanding of NP properties. On the experimental side, the main uncertainty is related to unknown value of external ligand/solvent dielectric constant. Indeed, significant energy shifts have been observed when changing the ligand/solvent for given NPs. While this uncertainty has important effect on the prediction of exciton binding energies, it affects much less the prediction of ground optical transitions, for which self-energies and increased electron-hole interaction nearly compensate each other. The predicted huge values of exciton binding energies can be tested experimentally by comparing 1-photon and 2-photon absorption spectra, respectively giving access to S and P exciton states. The more direct measurement based on 1-photon absorption spectroscopy of 1S and 2S exciton states is unfortunately hampered by the presence of the strong light-hole 1S exciton transition and the somewhat weaker 1S split-off exciton (see table III).
On the modeling side, it is noteworthy that better account for the interface between the semiconductor and the organic ligand may affect the bare single particle states by changing barrier height and band offsets. Agreement with experimental data suggests that the rather common assumption that ligands act as a large potential for nanocrystal electronic states is physically valid. We note that thanks to 2D translational invariance, NP would allow realistic first-principle calculations of the organic/inorganic interface. As for excitonic effects, the role of over-simplifying assumptions like the effective mass approach and piecewise constant dielectric function should be estimated. More fundamentally, we find that binding energies can exceed the energy separation between heavy and light holes. Such situation was previously investigated for quantum wells \cite{Bauer1} and can lead to further significant increase of the binding energy. However, we insist on the robustness of the evaluation for self-energies and exciton binding energies:  equivalent calculations in a continuous medium, effective mass approximation, roughly fitting the envelope of tight binding amplitudes with sinewaves give values quite similar to those in tables \ref{t:table2} and \ref{t:table3}. 
Calculation of the electron-hole exchange interaction, that leads to a splitting between bright ($J_z= \pm1$) and dark ($J_z=\pm2$) states of the exciton is beyond the scope of this paper. However, since it scales with the exciton binding energy, we may readily expect considerable enhancement of the exciton exchange splitting up to a few meV range. Turning to the effect of finite lateral size, NP excitons have a Bohr radius  ($< 1$nm) much smaller than typical NP lateral size (50 nm), while the latter remains small compared to emitted wavelength. This corresponds to the combined regimes of center-of-mass quantization and dipolar emission \cite{Andreani1999}. More precisely, in this regime exciton spectrum consists of nearly uncoupled exciton internal state and center-of-mass state. To first order, only ``$S, J_z=\pm1$" internal states combined with translational ground state are radiative, with a ``giant" oscillator strength proportional to NP surface. Indeed, short recombination times have been evidenced in the early experiments, but measured values are probably limited by the scattering of radiative exciton ground state into non-radiative excited states of the center-of-mass motion. Low temperature spectroscopic investigations of single nano-platelets are highly desirable in order to delineate the limits of nanoplatelets ideality, and evidence the possible existence of an optimal lateral size.

\section{conclusion}
We have shown that semiconductor nanoplatelets are rather original objects where completely stable excitons should exist at room temperature.  The huge value of exciton binding energies is \textit{governed} by the strong increase of electron-hole interaction due to image charge effects for such ultra-thin semiconductor layers placed in a small refractive index surrounding. Conversely, the bandgap for separate electron and hole is considerably increased due to repulsive self-interaction between carriers and their own dielectric images. The predicted robustness of these excitons, the associated large oscillator strength and the small ensemble-broadening suggests that NPs (possibly inserted in optical microcavities) could be  valuable objects to study Bosonic effects (multi-exciton states, condensates) at room temperature.\\
\section*{ACKNOWLEDGEMENTS}
This work was partly supported by Triangle de la Physique ``CAAS",  ANR  ``PEROCAI" and ``SNAP", and by RFBF grants. Al.L.E.
acknowledges financial support of the Office of Naval Research through the Naval Research Laboratory Basic Research Program.

\bibliography{NP_excitons_paper9}
\bibliographystyle{apsrev4-1}

\end{document}

%% file: Author_list.tex
%
\newcommand{\affA}{Laboratoire de Photonique et Nanostructures, CNRS, 91460 Marcoussis, France}
\newcommand{\affB}{A. M. Prokhorov General Physics Institute, RAS, Russia and Institut Pascal, PHOTON-N2, CNRS, Clermont-Ferrand, France}
\newcommand{\affC}{FOTON, Universit\'e Europ\'eenne de Bretagne, INSA and CNRS, 35708 Rennes, France}
\newcommand{\affD}{Ioffe Physical-Technical Institute, Russian Academy of Sciences, St Petersburg, Russia}
\newcommand{\affE}{Laboratoire de Physique et d'Etude des Mat\'eriaux, CNRS and ESPCI, 75005 Paris, France}
\newcommand{\affF}{Naval Research Laboratory, Washington, DC 20375, USA}

\author{R.~Benchamekh}\affiliation{\affA}
\author{N. A.~Gippius}\affiliation{\affB}
\author{J.~Even}\affiliation{\affC}
\author{M. O.~Nestoklon}\affiliation{\affD}\affiliation{\affA}
\author{J.-M.~Jancu}\affiliation{\affC}
\author{S. Ithurria}\affiliation{\affE}
\author{B.~Dubertret}\affiliation{\affE}
\author{Al. L.~Efros}\affiliation{\affF}
\author{P.~Voisin}\affiliation{\affA}

%
%
%
\vskip 0.25cm

%% file: NP_excitons_paper9.bbl
\begin{thebibliography}{29}
\bibitem [{not()}]{notewv}%
  \BibitemOpen
  \href@noop {} {}\bibinfo {note} {Note that, following tradition, what is
  plotted in the right panels is trajectories from $X$ to $U$ followed by
  trajectories from $K$ to $\Gamma$.}\BibitemShut {Stop}%
\end{thebibliography}
